\documentclass[aps,pra,twocolumn,showpacs,floatfix,amsmath,amssymb,nofootinbib,superscriptaddress]{revtex4-1} 

\usepackage{amsmath}
\usepackage{amsfonts}
\usepackage{graphicx}
\usepackage{bm}

\newcommand{\meq}{\mathrm{eq}}

\begin{document}

\title{Nonequilibrium scenarios in cluster-forming quantum lattice models}

\author{Adriano Angelone}
\affiliation{Abdus Salam ICTP, Strada Costiera 11, I-34151 Trieste, Italy}
\affiliation{SISSA, Via Bonomea 265, I-34136 Trieste, Italy}

\author{Tao Ying}
\affiliation{Department of Physics, Harbin Institute of Technology, 150001 Harbin, China}

\author{Fabio Mezzacapo}
\affiliation{Univ Lyon, ENS de Lyon, Univ Claude Bernard, CNRS, Laboratoire de
Physique, F-69342 Lyon, France}

\author{Guido Masella}
\affiliation{icFRC and ISIS (UMR 7006), Universit\'e de
Strasbourg and CNRS, 67000 Strasbourg, France}

\author{Marcello Dalmonte}
\affiliation{Abdus Salam ICTP, Strada Costiera 11, I-34151 Trieste, Italy}
\affiliation{SISSA, Via Bonomea 265, I-34136 Trieste, Italy}

\author{Guido Pupillo}
\affiliation{icFRC and ISIS (UMR 7006), Universit\'e de
Strasbourg and CNRS, 67000 Strasbourg, France}

\date{\today}

\begin{abstract}
  We investigate the out-of-equilibrium physics of monodisperse bosonic
  ensembles on a square lattice. The effective Hamiltonian description of these
  systems is given in terms of an extended Hubbard model with cluster-forming
  interactions relevant to experimental realizations with cold Rydberg-dressed
  atoms. The ground state of the model, recently investigated in Phys.  Rev.
  Lett. 123, 045301 (2019), features, aside from a superfluid and a stripe
  crystalline phase occurring at small and large interaction strength $V$,
  respectively, a rare first-order transition between an isotropic and an
  anisotropic stripe supersolid at intermediate $V$. By means of quantum Monte
  Carlo calculations we show that the equilibrium crystal may be turned into a
  glass by simulated temperature quenches and that out-of-equilibrium isotropic
  (super)solid states may emerge also when their equilibrium counterparts are
  anisotropic. These out-of-equilibrium states are of experimental interest,
  their excess energy with respect to the ground state being within the energy
  window typically accessed in cold atom experiments. We find, after quenching,
  no evidence of coexistence between superfluid and glassy behavior. Such an
  absence of superglassiness is qualitatively explained.
\end{abstract}

\maketitle

\section{Introduction}

The search for ordered or disordered exotic states of matter is a very active
field of investigation in condensed matter physics \cite{Boninsegni2012-2,
Nandkishore2015, Qi2010}. The interactions between the individual constituents
of a given system play a fundamental role in this context, being intrinsically
related to the physical mechanisms responsible for the stabilization of
different (possibly novel) physical scenarios. Usually, intriguing equilibrium
or out-of-equilibrium (OOE) properties emerge in the presence of frustration,
i.e., the impossibility of simultaneously satisfying a minimum energy condition
for all terms of the Hamiltonian (see, e.g., Refs. \cite{Edwards1975,
Binder1986}). The latter may arise from, e.g., competing interactions, the
presence of peculiar substrates (i.e., lattices) or polidispersity, i.e., the
presence in the system of particles with different properties such as, for
example, mass and/or size.

\begin{figure}
  \centerline{\includegraphics*[width=0.90\columnwidth]{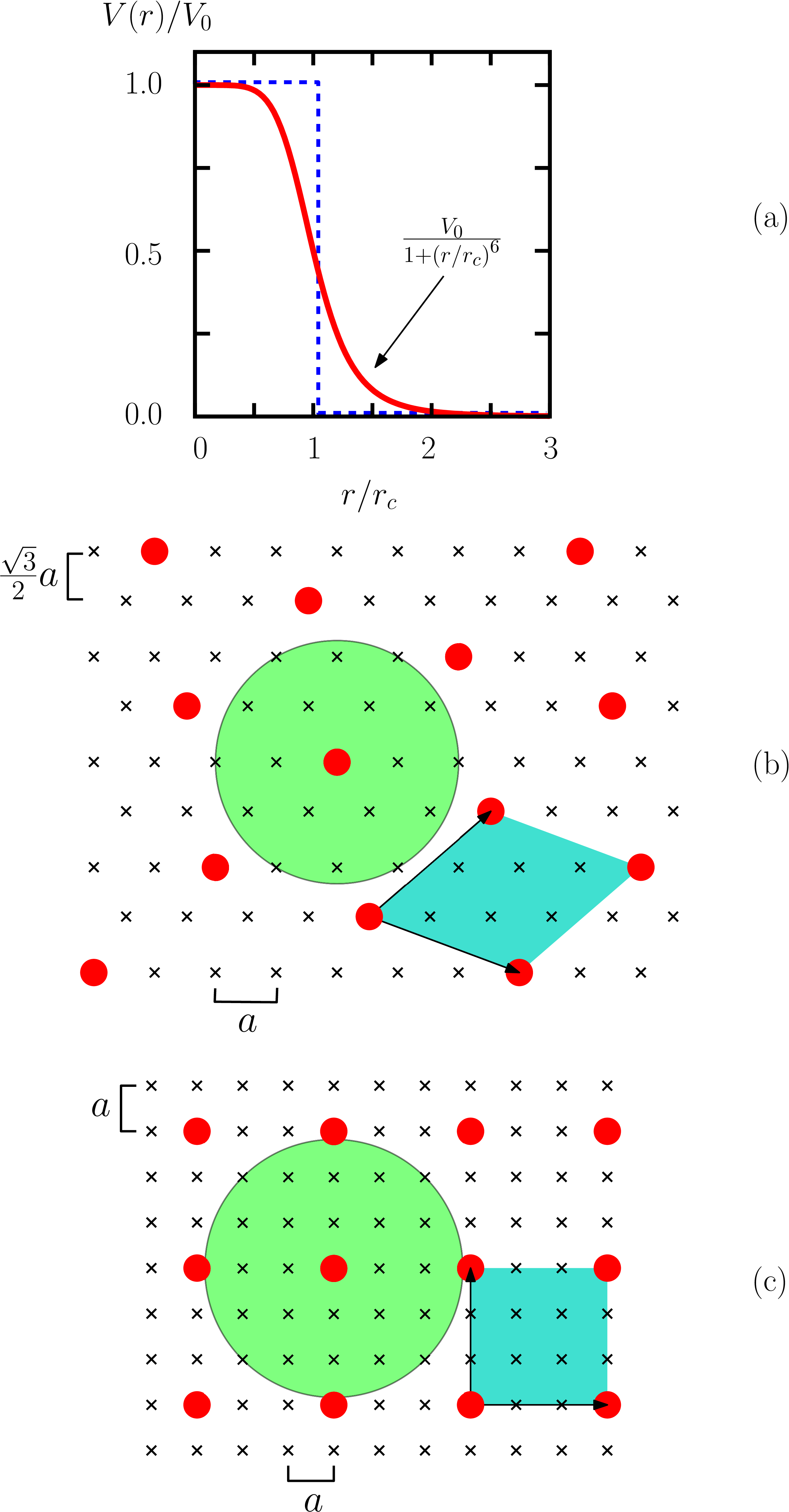}}
  \caption{Panel (a): examples of extended-range interactions. The continuous
  and dashed curves represent a soft-shoulder potential (see label) with
  fast-decaying [i.e., $(r/r_c)^{-6}$] tail and a shoulder interaction of
  strength $V_0$ and radius $r_c$, respectively. Single-particle crystalline
  ground states of the extended Bose-Hubbard Hamiltonian in Eq.
  \eqref{eq.hamiltonian} on a triangular lattice for $r_c = 2a$ and $\rho =
  1/7$ [panel (b)], and on a square lattice for $r_c = 2 \sqrt{2} a$ and $\rho
  = 1/9$ [panel (c)]. In panels (b) and (c), black crosses are lattice sites
  and red dots are occupied sites. Green circles highlight the range of the
  interaction around an occupied site, black arrows are the vectors generating
  the crystalline structure, and cyan regions indicate the primitive cells of
  the crystals.}
  \label{FigCrystals}
\end{figure}

Recently, a large class of purely repulsive, isotropic extended-range
interactions (ERI), whose relevance ranges from classical soft-matter
systems~\cite{Mladek2006, Lenz2012, Sciortino2013} to cold Rydberg-atom
experiments~\cite{Saffman2010, Jau2015, Zeiher2016, Lahaye2009, Low:2012aa,
Bernien:2017aa, Leseleuc:aa}, has elicited considerable theoretical interest.
Indeed, these potentials offer the possibility to explore a variety of
equilibrium and OOE phenomena in realistic models where frustration, in the
forms discussed above, is not included. The main features of pairwise ERI are a
plateau which extends up to interparticle distances of the order of the
critical radius $r_c$ and a tail quickly approaching zero for $r > r_c$ [see
Fig.~\ref{FigCrystals}(a)]. Systems with ERI at high enough particle density
$\rho$ are characterized, in the classical limit, by a so-called cluster
crystalline ground state (GS) where crystalline sites are occupied by
self-assembled aggregates of particles (i.e., clusters). Classical cluster
crystals have been shown to possess peculiar equilibrium dynamical properties
resembling those of glass-forming liquids, while still retaining structural
order \cite{Diaz-Mendez2015}. For these systems, OOE glassy scenarios where
disorder coexists with clusterization have also been
predicted~\cite{Diaz-Mendez2017, Diaz-Mendez2019}. Furthermore, when quantum
effects are taken into account, clusterization may lead to anomalous
Luttinger-liquid behavior in one spatial dimension (1D) \cite{Mattioli2013,
Dalmonte:2015aa, Motta:2016aa, Teruzzi:aa}, as well as to the coexistence of
diagonal long range order and superfluidity (i.e., supersolidity) in 2D free
space~\cite{Henkel2010, Henkel2012, Cinti2010, Cinti2014} or on a triangular
lattice~\cite{Angelone2016}. In the latter, superfluidity may also be
concomitant to glassiness in a so-called OOE superglass.

In order to gain theoretical insight into the novel physical phenomena related
to clusterization, as well as into its interplay with quantum effects, system
geometry, and interaction radius $r_c$, it is of crucial interest to extend the
investigation to different lattices and choices of relevant parameters. In this
context, a recent work by some of us \cite{Masella2019} has been devoted to the
study of the GS phase diagram of a cluster-forming model of hard-core bosons
with shoulder ERI on a square lattice. For such a model the GS is a
superfluid (stripe crystal) for sufficiently small (large) interaction strength
$V$. Surprisingly, for intermediate values of $V$ a first-order phase
transition occurs between two different supersolids: an isotropic one, emerging
from the superfluid when $V$ is increased, and an anisotropic stripe supersolid
emerging from the partial quantum melting of the large-$V$, essentially
classical, crystal.

The study of the GS phases mentioned above required extensive calculations and
careful temperature and interaction annealings due to the presence of many OOE
states close in energy to the GS. Indeed, it is known that extended-range
interactions on a lattice lead to a plethora of low-energy metastable states,
whose number exponentially increases with the system size~\cite{Menotti2007,
Trefzger2008}. Such OOE states are of importance to possible quantum
simulations of our model of interest with cold atoms since in these experiments
states whose energy is above the GS one are commonly accessed (even when one is
interested in GS physics, due in this case to the presence of undesired
excitations). As an example, even in the case of the preparation of a bosonic
Mott insulator (a comparatively simpler state with respect to the ones
discussed in this work) one can currently obtain up to defect densities of up
to the percent level, which (in the strongly interacting regime) correspond to
differences in energy density of up to a few percent. As a consequence, the
fundamental question which need be addressed is whether or not theoretical
predictions that one can make for the GS are stable within the characteristic
energy window accessible to experiments~\cite{Bloch2012}.

In this work the OOE scenarios of the model studied in~\cite{Masella2019} are
systematically investigated by means of a path integral Monte Carlo (PIMC)
approach. In particular, our system can be driven out of equilibrium via PIMC
low-temperature ($T$) quenches.

Our main findings are the following. (i) As opposed to the isotropic and
anisotropic supersolid GSs, a low-temperature quench leads to largely
isotropic, OOE (super)solid states. Remarkably, these are also found for
values of $V$ at which the equilibrium phases are instead anisotropic. (ii)
Similarly to our previous study of the same (albeit with different $r_c$ and
$\rho$) quantum model on the triangular lattice \cite{Angelone2016}, as well
as to that of the classical model in free space \cite{Diaz-Mendez2017}, the
OOE counterpart of the equilibrium crystal at large $V$ is a normal glass.
(iii) In the investigated parameter range no evidence of superglassy behavior
is obtained. The occurrence of such a state, which has been predicted for the
triangular lattice, crucially depends on the interplay between lattice
geometry, particle density, and interparticle interactions.

It is worth mentioning that the energy deviations from the GS of the OOE states
analyzed in this work are comparable to those routinely obtained in cold atom
experiments.

The remainder of this paper is organized as follows. In the next section we
describe the details of the Hamiltonian model of our interest with particular
attention to its cluster-forming regimes and, briefly, the numerical method
adopted to carry out our investigation. In Sec. III we present and discuss
our results, while in the last section we outline the conclusions of our work.

\section{Model and Methods}

The model we investigate is described by the Hamiltonian

\begin{equation}    \label{eq.hamiltonian}
H = - t \sum_{\langle ij \rangle} \left( b_i^{\dagger} b_j + \mathrm{h.c.}
\right) + V \!\!\!\! \sum_{i < j : r_{ij} \leq r_c} n_i n_j
\end{equation}

on a square lattice of $N = L \times L$ sites and lattice constant $a$ with
periodic boundary conditions. Here $t$ is the hopping coefficient between
nearest-neighbor sites, $b_i$ and $b_i^{\dagger}$ are annihilation and creation
operators for hard-core bosons on site $i$, respectively, $n_i = b^{\dagger}_i
b_i$, $V$ is the interaction strength, and $r_{ij}$ is the distance between
sites $i$ and $j$. In the following, $a$ and $t$ will be taken as units of
length and energy, respectively.

For $r_c = a$, i.e., nearest-neighbor potential, the phase diagram of the model
contains superfluid, solid, and insulating phases \cite{Schmid2002}, while
supersolid states can be stabilized adding longer-ranged density-density
interactions \cite{Batrouni2000, Hebert2002}. We study the model for $r_c > a$,
in a regime where cluster formation takes place in the system.

For low enough $\rho$, the classical (i.e., $t = 0$) GS is a zero-energy
single-particle crystal, where the interparticle spacing is larger than the
interaction radius. The maximum density $\rho_c$ for which such a crystal
exists is determined by $r_c$ and the lattice geometry. For example, in our
study of Eq.~\eqref{eq.hamiltonian} on the triangular lattice the choice $r_c =
2a$ results in a critical density $\rho_c^{\mathrm{tr}} = 1/7$. In the
equilibrium study performed in~\cite{Masella2019}, on the other hand, a square
lattice geometry with $r_c = 2 \sqrt{2} a$ leads to a critical density
$\rho_c^{\mathrm{sq}} = 1/9$. The single-particle crystalline structures
corresponding to these densities are shown in Figs.~\ref{FigCrystals}(b)
and~\ref{FigCrystals}(c), respectively.

\begin{figure*}
  \centerline{\includegraphics*[width=0.80\textwidth]{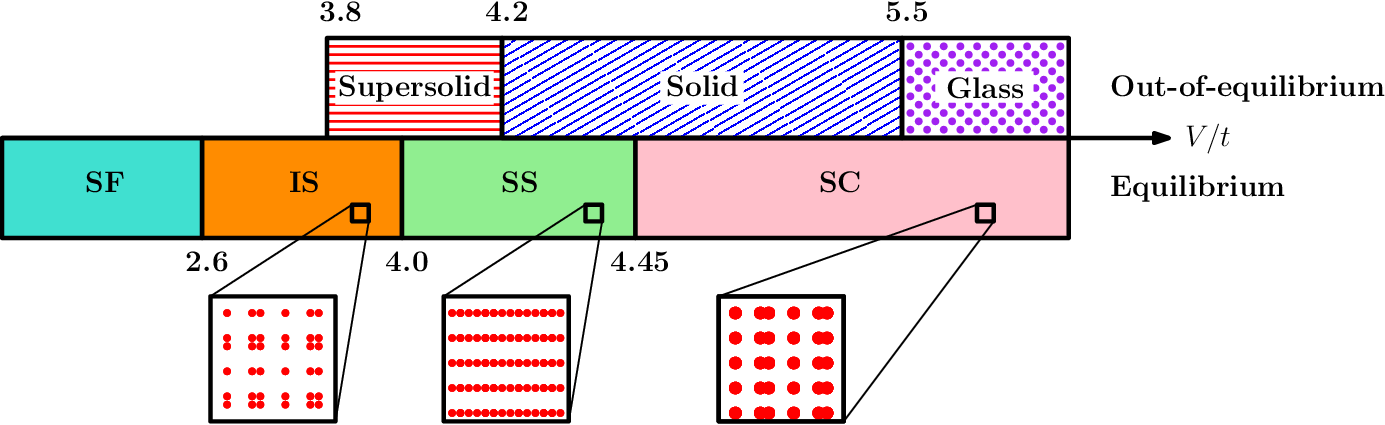}}
  \caption{Schematic phase diagram of Eq.~\eqref{eq.hamiltonian} as a function
  of the interaction strength $V/t$. Each colored region in the lower part of
  the figure corresponds to a GS equilibrium phase: namely, a superfluid (SF,
  cyan), an isotropic supersolid (IS, orange), a stripe supersolid (SS, green)
  and a stripe crystal (SC, pink). The drawings are sketches of the crystalline
  structure (where present) of each equilibrium phase. The filling patterns in
  the upper part of the diagram identify the OOE states reached via simulated
  temperature quenching at target temperature $T/t = 1/20$. The regions where
  quenching leads to OOE supersolid, solid and glassy states are denoted by
  horizontal, diagonal, or dot filling patterns, respectively.}
  \label{FigPhaseDiagram}
\end{figure*}

For $\rho \gtrsim \rho_c$, a single-particle solid has a higher potential
energy than a solid in which particles group up in tightly packed clusters.
Indeed, the latter can arrange themselves far enough from each other to be
noninteracting (i.e., outside of their mutual interaction radius). A larger
value of $\rho$ results in the formation of larger clusters. For instance, on
the triangular lattice the chosen value of $\rho = 13/36 \sim 2.5
\rho_c^{\mathrm{tr}}$ led to clusters of three to four particles on average,
while on the square lattice, for $\rho = 5/36 \sim 1.25 \rho_c^{\mathrm{sq}}$,
the largest clusters contain two particles.

When the system is driven away from thermal equilibrium, cluster formation can
cause effective polidispersity, which in turn plays a fundamental role in the
appearance of (super)glassy states~\cite{Angelone2016, Diaz-Mendez2017}. This
phenomenon is favored by large cluster sizes, as well as (sufficiently) strong
interactions, which prevent particles from delocalizing between different
clusters, and entire clusters from spatially rearranging to establish an
ordered (crystalline) structure.

In this work we analyze the OOE physics of Eq.~\eqref{eq.hamiltonian} for the
same parameter range investigated in~\cite{Masella2019}. The model shows a rich
GS phase diagram, characterized by, e.g., competing supersolid phases.
Understanding the robustness of these scenarios within the typical experimental
accuracy is one of the main objectives of the present study. Furthermore here,
due to the presence of significantly smaller clusters, frustration effects
should be significantly weaker than those occurring in the study of
Ref.~\cite{Angelone2016} on the triangular lattice. This would allow one to
determine, for instance, to which degree various OOE phenomena depend on
clusterization.

We study the model Eq.~\eqref{eq.hamiltonian} by means of path integral quantum
Monte Carlo simulations using Worm updates~\cite{Prokofev1998-1}. This is a
state-of-the-art technique, which yields numerically exact results for
unfrustrated bosonic systems and allows one to accurately estimate observables such
as the superfluid fraction $\rho_s/\rho = \left( 4 \beta t \rho \right)^{-1}
\langle W_x^2 + W_y^2 \rangle$ and the static structure factor $S(\mathbf{k}) =
N^{-2} \sum_{ij} \exp \left[ -i \mathbf{k} \left( \mathbf{r}_i - \mathbf{r}_j
\right) \right] \langle n_i n_j \rangle$. These order parameters measure
superfluidity and crystalline order, respectively, and are defined in terms of
the inverse temperature $\beta = (k_B T)^{-1}$ ($k_B$ is the Boltzmann
constant, set to one in the following), of the winding number $W_x, W_y$ in
direction $x,y$, respectively, and of the lattice wave vectors $\mathbf{k}$.
Here, $\langle \ldots \rangle$ stands for statistical average. We also estimate
the renormalized Edwards-Anderson parameter $Q_{\mathrm{EA}} = \sum_i \langle
n_i - \rho \rangle^2 / Q_{\mathrm{EA}}^0$, a well-known observable which allows one
to identify glass behavior in lattice systems in the absence of crystalline
order. The normalization $Q_{\mathrm{EA}}^0 = N \rho (1 - \rho)$ is the value
obtained for a fully localized state. Finally, we determine the single-particle
Green function defined as $G(\mathbf{r}) = N^{-1} \sum_i \langle b^{\dagger}_i
b_{i + \mathbf{r}} \rangle$, associated to the presence of off-diagonal
quasi-long-range order in our two-dimensional system.

We perform large-scale simulations with up to $N = 96 \times 96$ sites and
temperatures between $T/t = 1$ and $T/t = 1/20$, the latter yielding
essentially GS results in the equilibrium case~\cite{Masella2019}. To gain
insight into the OOE scenarios, we employ a simulated quench protocol, by
running low-$T$ simulations starting from high-$T$ configurations without
performing simulated annealing steps in $T$. The experimental relevance (in the
sense discussed in Sec. I) of our obtained OOE states is assessed \textit{a
posteriori}. In particular, our estimated OOE energies never exceed the GS ones
by more than 3.5\%.

\section{Results}

For clarity, we begin our discussion by summarizing the GS phase diagram of
model Eq.~\eqref{eq.hamiltonian} (we refer the reader to
Ref.~\cite{Masella2019} for an exhaustive discussion). The GS (lower part of
Fig.~\ref{FigPhaseDiagram}) is a superfluid (SF) at weak interactions, which
makes way for an isotropic supersolid (IS) at $V/t = 2.6$.  The system then
undergoes a first-order transition at $V/t = 4.0$ to a supersolid state with
anisotropic stripe crystalline structure and superfluid response, i.e., a
stripe supersolid (SS). Finally, superfluidity is lost at $V/t = 4.45$, and the
GS becomes a stripe crystal (SC).\\

\begin{figure}[h!]
  \centerline{\includegraphics*[width=0.90\columnwidth]{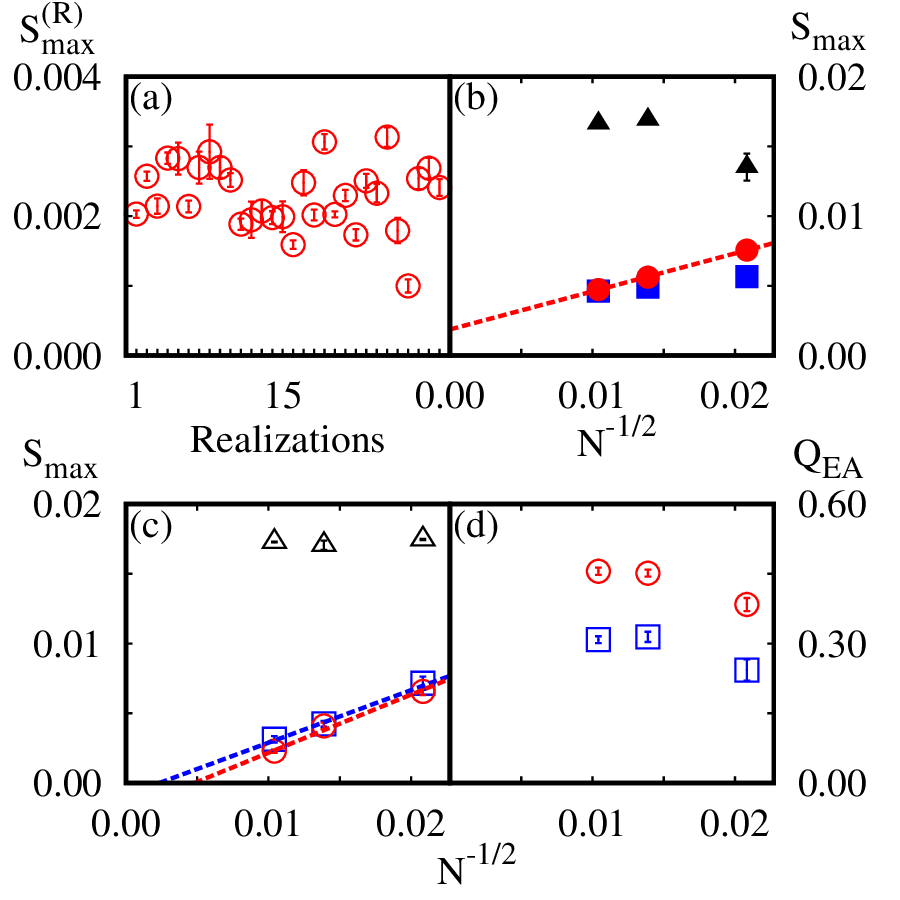}}
  \caption{Results for finite-temperature simulated quenches performed at $V/t
  = 5.0$ and $V/t = 6.0$. Panel (a): maximum value of the realization-dependent structure factor
  $S^{(R)}_{\max}$ as a function of the realization index for $L = 96$, $T/t =
  1/20$, and $V/t = 6.0$. Panel (b): realization-averaged value of $S_{\max}$
  as a function of the inverse system size for $V/t = 5.0$. Panel (c): same as
  for panel (b) for $V/t = 6.0$. Panel (d): realization-averaged Edwards-Anderson parameter
  $Q_{\mathrm{EA}}$ as a function of the inverse system size for $V/t = 6.0$.
  In all panels, filled (empty) symbols correspond to $V/t = 5.0$ ($V/t =
  6.0$), while triangles, squares and circles correspond to $T/t = 1/5, 1/10,
  1/20$, respectively. The dashed lines correspond to linear fits in
  $N^{-1/2}$, shown when estimates for the two largest sizes are not identical
  within numerical uncertainty.}
  \label{FigHighV}
\end{figure}

\begin{figure}[h!]
  \centerline{\includegraphics*[width=0.90\columnwidth]{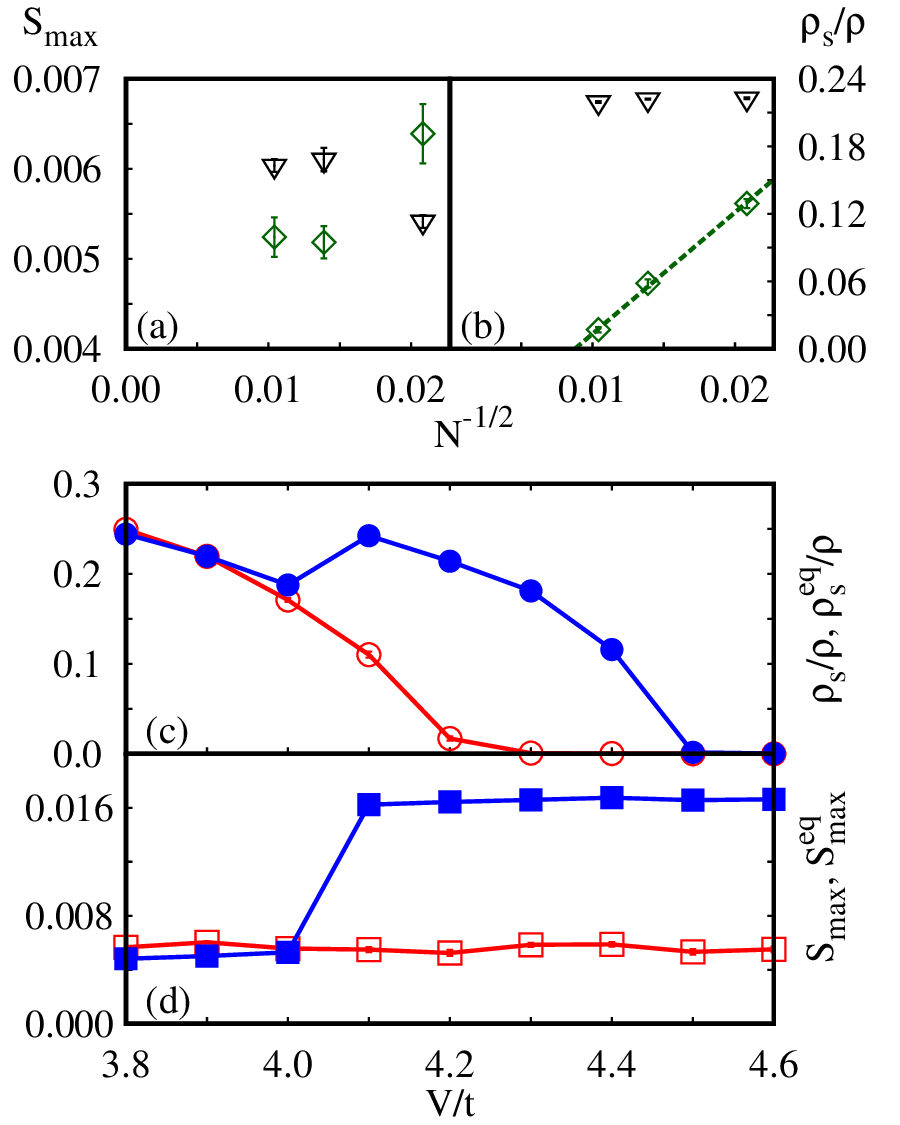}}
  \caption{Panels (a),(b): scaling in the inverse size of $S_{\max}$ and
  $\rho_s/\rho$ for $V/t = 3.9$ (triangles) and $V/t = 4.2$ (diamonds) at $T/t
  = 1/20$. Dashed lines are linear fits to the numerical data, shown when
  estimates for the two largest sizes are not identical within their
  uncertainty. Panel (c): comparison between the equilibrium superfluid
  fraction $\rho_s^{\meq}/\rho$ (filled circles) and the OOE one
  $\rho_s/\rho$ (empty circles) as a function of the interaction strength
  at $T/t = 1/20$ and $L = 96$. Panel (d): comparison of the equilibrium
  maximum value of the structure factor $S^{\meq}_{\max}$ (filled
  squares) and the OOE one $S_{\max}$ (empty squares) for the same
  parameters of panel (c). In panels (c) and (d), solid lines are guides to the
  eye. In all panels, OOE estimates are realization-averaged.}
  \label{FigMediumV}
\end{figure}

Driving the system away from thermal equilibrium results in the OOE phase
diagram shown in the upper part of Fig.~\ref{FigPhaseDiagram}, obtained via
analysis of the relevant observables shown in Figs.~\ref{FigHighV} and
\ref{FigMediumV} and discussed below. In the strongly interacting regime, i.e.,
$V/t > 5.5$, and for temperatures $T/t < 1/5$, the simulated quenches stabilize
OOE states where diagonal long range order vanishes in the thermodynamic limit.
As signaled by the finite value of $Q_{\mathrm{EA}}$ (see Fig.~\ref{FigHighV}),
concomitant to the absence of superfluidity, the resulting states are normal,
essentially classical, glasses \cite{questionB-ref1}.

Conversely, following our quenches at $T/t = 1/20$ in the intermediate-$V/t$
region (i.e., $3.8 \leq V/t \leq 4.6$) the system retains long-range order,
reaching OOE states with crystalline structures different from those obtained
at equilibrium [see Fig.~\ref{FigMediumV}(d)]. These results allow one to identify
a variety of crystalline states in the OOE phase diagram, as
different realizations of each simulation may converge to states with different
strength and type of diagonal long-range order. For $V/t < 4.2$ the system
displays superfluid behavior [see Fig.~\ref{FigMediumV}(c)]. The latter
coexists with diagonal long-range order down to $V/t = 3.8$, pointing out the
occurrence of OOE supersolid states in this parameter range. For $V/t < 3.8$
our quenching process is ineffective, and the system equilibrates to an IS and
a SF for $V/t > 2.6$ and $V/t < 2.6$, respectively. Remarkably, the OOE
supersolids display features considerably different from their equilibrium
counterparts. Specifically, both superfluid responses and crystalline order are
essentially isotropic even when the corresponding equilibrium supersolids are
strongly anisotropic.

In both the high- and intermediate-$V/t$ region, we determine the degree of
equilibration of each simulated quench by performing it in several (i.e.,
$\gtrsim 30$) independent realizations, differing in both the initial
configuration and in the thermalization seed of the QMC simulation.
This procedure mimics what is customarily done in the laboratory,
where many realizations of an experiment are performed to measure the
observables of interest.

If our quench protocol does not drive the system away from thermal equilibrium,
all the realizations converge to the equilibrium state, and the details of the
QMC stochastic dynamics in configuration space are inessential. On the other
hand, where the OOE driving succeeds, most of the realizations fail to
equilibrate, with their initial conditions becoming crucial in determining the
state reached by each simulation. A typical example of the latter behavior is
shown in Fig.~\ref{FigHighV}(a), where the mean value of an observable (in this
case, the maximum value of the structure factor) strongly depends on the
realization.

In general, for different realizations of a quench at a chosen
parameter set, observables may take different values; however, each realization
is found to display the hallmark features of the corresponding OOE phase. For
example, in the OOE supersolid region of Fig.~\ref{FigPhaseDiagram} some
realizations may yield crystalline orders of different strength, as well as
different values of superfluid fraction. Nevertheless, each realization
displays finite superfluid fraction and concomitant diagonal long-range order.

Figures~\ref{FigHighV}(b)-\ref{FigHighV}(d) show the scaling, as a function of
the system size, of the realization-averaged maximum peak of the structure
factor $S_{\max}$ and of the Edwards-Anderson parameter $Q_{\mathrm{EA}}$ after
quenching to different target temperatures for $V/t = 5.0$ and $V/t = 6.0$
(filled and empty symbols, respectively). For these values of $V/t$, the
observed OOE states are non-superfluid. We find equilibration to a stripe solid
for $T/t = 1/5$ (triangles). A decrease of the target temperature results in
failure to equilibrate the vast majority of realizations, which converge to
states where diagonal long range order is suppressed with respect to the
equilibrium scenario. For $V/t = 5.0$, $S_{\max}$ remains finite in the
thermodynamic limit for $T/t = 1/10$ (filled squares) and $T/t = 1/20$ (filled
circles), signaling an OOE crystal. Conversely, crystalline order is lost for
$V/t = 6.0$ and $T/t = 1/10$ (empty squares) and $T/t = 1/20$ (empty circles).
For these temperatures, $Q_{\mathrm{EA}}$ remains finite [panel (d)], signaling
the emergence of glassy behavior.

Figure~\ref{FigMediumV} shows a detailed comparison of the realization-averaged superfluid and
crystalline order parameters for the OOE and equilibrium cases (empty and filled
symbols, respectively) as a function of $V/t$. Supersolid behavior occurs,
after quenching, for $3.8 \leq V/t < 4.2$, i.e., in an interaction strength
window smaller than that for which supersolidity is found at equilibrium. In
particular, superfluidity vanishes in the thermodynamic limit for $V/t \geq
4.2$ [diamonds in Fig.~\ref{FigMediumV}(b)]. As mentioned above, the details of
these supersolid OOE states may be both quantitatively and qualitatively
different from the equilibrium ones, whose order parameters are denoted for
clarity by $\rho_s^{\meq}/\rho$ and $S_{\max}^{\meq}$, respectively. As
expected, for small $V/t \lesssim 4$ our quenching protocol does not
significantly alter the values of $\rho_s/\rho$ and $S_{\max}$ with respect to
$\rho_s^{\meq}/\rho$ and $S_{\max}^{\meq}$ (Fig.~\ref{FigMediumV}(c) and~\ref{FigMediumV}(d) and
finite-size scaling for $V/t = 3.9$ [triangles in panels (a) and (b)]).  For
intermediate $V/t$, while at the IS/SS transition $S_{\max}^{\meq}$
[filled squares in Fig.~\ref{FigMediumV}(d)] develops strong
anisotropy~\cite{Masella2019} and features a sizable variation, $S_{\max}$
remains essentially constant [empty squares in
Fig.~\ref{FigMediumV}(d)] and isotropic. Indeed, in all quench realizations the
maximum peaks of the structure factor occur at realization-dependent
wave vectors $\left( k_x^{(R)}, 0 \right)$ and $\left( 0, k^{(R)}_y \right)$
with $k^{(R)}_x \simeq k^{(R)}_y$ and $S \left( k_x^{(R)}, 0 \right)
\simeq S \left( 0, k_y^{(R)} \right)$. Similarly, $\rho_s/\rho$
[empty circles in Fig.~\ref{FigMediumV}(c)] takes considerably lower
values than $\rho_s^{\meq}/\rho$ [filled circles in
Fig.~\ref{FigMediumV}(c)] for $V/t \gtrsim 4.0$ and the superfluid response, as
opposed to what is found at equilibrium, is essentially isotropic. This clarifies
the difference between the equilibrium supersolid states, which can be either
isotropic or anisotropic, and the OOE ones, which are found to be always
largely isotropic. Such a difference persists even in the absence of
superfluidity in the OOE states: for example, for $V/t > 4.45$ the GS is a
stripe crystal while quenching results in the appearance of substantially
isotropic crystals and glasses.

\begin{figure}[h!]
  \centerline{\includegraphics*[width=0.90\columnwidth]{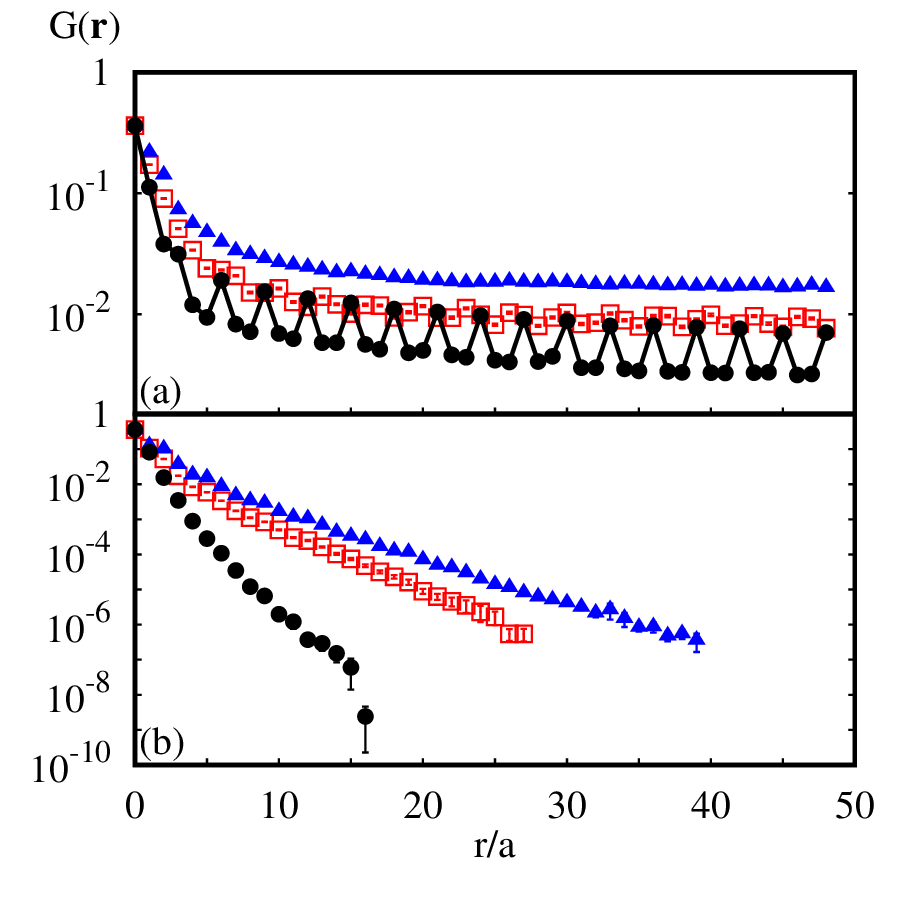}}
  \caption{Panel (a): single-particle Green function $G(\mathbf{r})$ for $T/t =
  1/20$, $L = 96$, and $V/t = 4.1$. Triangles and circles refer to the
  equilibrium $G(\mathbf{r})$ along the $x$ and $y$ direction, respectively
  (see text), while squares denote the OOE $G(\mathbf{r})$ along the $y$
  direction. The corresponding OOE $G(\mathbf{r})$ along the $x$ direction (not
  shown) is essentially identical. Continuous lines are guides to the eye.
  Panel (b): same as panel (a) for $V/t = 6.0$.}
  \label{FigGreen}
\end{figure}

The isotropic character of the OOE states can also be inferred by inspection
of $G_x$ and $G_y$, i.e., the single-particle Green function $G(\mathbf{r})$
along the $x$ and $y$ directions, respectively. For $V/t = 4.1$, $G_x$ and
$G_y$ of the corresponding anisotropic SS GS [triangles and circles in
Fig.~\ref{FigGreen}(a), respectively] are clearly different. Specifically,
while both decay algebraically as a function of the distance, signaling
off-diagonal quasi-long-range order, $G_y$ is characterized by oscillations in
correspondence of the stripe periodicity. The OOE $G(\mathbf{r})$, on the
other hand, is essentially isotropic, i.e., $G_x \sim G_y$ [squares in
Fig.~\ref{FigGreen}(a)].

Figure~\ref{FigGreen}(b) displays the same comparison for $V/t = 6.0$. Here
the decay of the $G(\mathbf{r})$, both at equilibrium and OOE, is exponential,
as expected for a nearly classical crystal and a glass, respectively. Also in
this case, the equilibrium $G(\mathbf{r})$ is strongly anisotropic, while $G_x
\sim G_y$ in its OOE counterpart.

\begin{figure}
  \centerline{\includegraphics*[width=0.85\columnwidth]{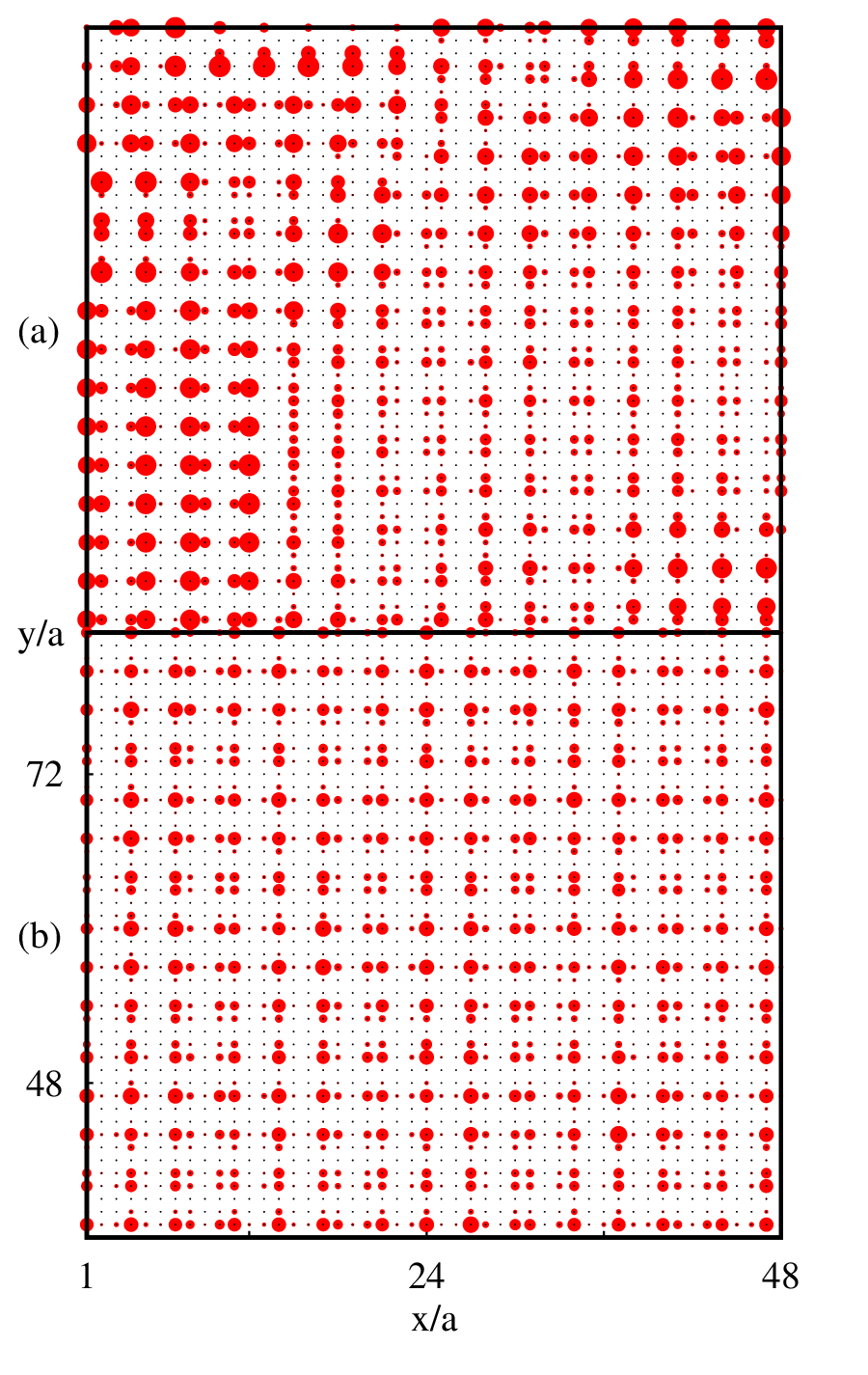}}
  \caption{Panel (a): portion of a site occupation map for one realization for
  $L = 96$, $T/t = 1/20$, and $V/t = 6.0$. Panel (b): same as panel (a) for
  $V/t = 4.1$. In both panels, black dots correspond to lattice sites, while
  the size of the red dot on each site is proportional to its occupation.}
  \label{FigSnapshots}
\end{figure}

Further insight into the OOE physics of our model can be gained from the
occupation maps in Fig.~\ref{FigSnapshots}. In both cases shown in the figure
[$V/t = 6.0$ in panel (a) and $V/t = 4.1$ in panel (b)] particles clusterize;
for $V/t = 6.0$, clusters have in general different shapes and orientations.
These induce an effective polidispersity, ultimately resulting in glassy
behavior~\cite{Diaz-Mendez2017}. On the other hand, for $V/t = 4.1$, where the
system is supersolid, particles can ``hop'' between different clusters,
establishing long exchange cycles which give rise to a sizable superfluid
response. The latter is concomitant with a well defined crystalline structure.
While different kinds of diagonal long-range order may appear depending on the
realization, the vast majority of the latter lead to the same type of
crystalline order of the supersolid state shown in Fig. 6(b).

It is important to mention that no evidence of superglassy behavior has been
found in the parameter range investigated in this work. This constitutes an
important difference with respect to the results of our study of model
Eq.~\eqref{eq.hamiltonian} for $r_c = 2$ on the triangular
lattice~\cite{Angelone2016}. Qualitative insight into this
difference can be obtained by analyzing the two fundamental ingredients of a
superglass phase: namely, superfluidity and frustration.

A source of frustration, in our observed OOE states, is cluster formation, and
the resulting self-induced effective polidispersity \cite{Diaz-Mendez2017}. In
this framework, frustration is magnified for increasing cluster sizes (whose
typical value is related to the ratio $\rho/\rho_c$, see Sec. II) and/or
interaction strength, since cluster formation takes place as a mechanism to
decrease the potential energy of the system, and this effect is more important
when $V/t$ is large. This allows one to have strong frustration even with very
small clusters, but in general may not allow one to obtain a superglass, since
for very strong interactions superfluidity is also suppressed, and one obtains
only a normal (i.e., insulating) glass.

In our study on the triangular lattice we choose $\rho \sim 2.5
\rho_c^{\mathrm{tr}}$; as a result, clusters of three to four particles are formed
in the obtained OOE states. Superglassy behavior is found because
frustration-induced suppression of crystalline order occurs, due to the
relatively large cluster size, at moderate values of $V/t$, where a sizable
superfluid fraction is found at equilibrium and survives in the OOE scenario.
Conversely, on the square lattice we consider $\rho \sim 1.25
\rho_c^{\mathrm{sq}}$. As discussed above, this results in smaller clusters and
weaker frustration effects at a given $V/t$. In order to observe the structural
order loss characteristic of a glassy state one therefore needs to go at
interaction strengths high enough ($V/t \sim 5.5$) that superfluidity is
suppressed.

A further numerical indicator of the difference between the two systems can be
given using a simple energetic argument. Indeed, the ratio $\Lambda$ between
potential and kinetic energy can be used to roughly estimate particle mobility.
When $\Lambda$ is large, e.g., for large $V/t$ or large $\rho/\rho_c$,
particles, and clusters formed after quenching, are strongly localized,
preventing the realization of a crystalline structure. Superglassy behavior
emerges as a delicate balance between localization and superfluidity, which
conversely takes place, at low $T/t$, for small $\Lambda$.

Indeed, on the triangular lattice superglasses were observed as the OOE
counterparts of supersolids at $\Lambda \sim 9$ and $\rho_s/\rho \sim 0.1$. On
the other hand, the equilibrium supersolid phases of model
Eq.~\eqref{eq.hamiltonian} on the square lattice are characterized by much
smaller $\Lambda \sim 1.2$ and higher $\rho_s/\rho \sim 0.2$. Here the OOE
driving leads to superfluid states where, due to larger mobility, crystalline
order can be restored, and glassy states can only be obtained at larger $V/t$,
where superfluidity is suppressed.

\section{Conclusions and Outlook}

We study the out-of-equilibrium scenarios of a model of monodispersed hard-core
bosons on a square lattice with an extended-range potential of the
shoulder type, of interest for experiments with cold
Rydberg-dressed atoms. In the parameter region of our investigation, the ground
state of the model is a cluster crystal and a superfluid for strong and weak
interactions respectively, and a rare transition between an isotropic and an
anisotropic supersolid state occurs for intermediate interaction strength.

Via simulated temperature quenches, we obtain glasses for strong interactions,
while for moderate values of the latter (super)solids appear. Such
(super)solids are qualitatively different from the ground state ones, being
essentially isotropic even for values of the interaction strength for which the
corresponding ground state is anisotropic. The out-of-equilibrium states we
find display energy deviations with respect to the ground state which are
comparable with those of common cold atom experiments.

For all interaction strength values where out-of-equilibrium superfluidity
remains finite, long-range crystalline order is also maintained. Therefore, no
evidence of superglassy behavior is found in our region of investigation, as
opposed to the case of the triangular lattice, where we demonstrated such an
exotic state. Indeed, the choice of a smaller particle density on the square
lattice leads to frustration effects strong enough to cause the loss of
crystalline order only at an interaction value where superfluidity is also
lost.\\
\section{Acknowledgements}

We acknowledge useful discussion with A. Kuklov, W. Lechner, M. Mattioli, and
S. Wessel. Work in Strasbourg was supported by the grant ANR-ERA-NET QuantERA
-- Projet RouTe (No. ANR-18-QUAN-0005-01). G. P. acknowledges support from the
Institut Universitaire de France (IUF) and USIAS. G. M. was also supported by
the French National Research Agency (ANR) through the Programme
d'Investissement d'Avenir under contract ANR-17-EURE-0024. T. Y. was also
supported by the National Natural Science Foundation of China (Grant No.
11504067).  Work in Trieste was supported by the ERC under Grant No. 758329
(AGEnTh) and by the EU Quantum Flagship grant PASQuanS.

\bibliography{index.bib}

\begin{thebibliography}{36}%
\makeatletter
\providecommand \@ifxundefined [1]{%
 \@ifx{#1\undefined}
}%
\providecommand \@ifnum [1]{%
 \ifnum #1\expandafter \@firstoftwo
 \else \expandafter \@secondoftwo
 \fi
}%
\providecommand \@ifx [1]{%
 \ifx #1\expandafter \@firstoftwo
 \else \expandafter \@secondoftwo
 \fi
}%
\providecommand \natexlab [1]{#1}%
\providecommand \enquote  [1]{``#1''}%
\providecommand \bibnamefont  [1]{#1}%
\providecommand \bibfnamefont [1]{#1}%
\providecommand \citenamefont [1]{#1}%
\providecommand \href@noop [0]{\@secondoftwo}%
\providecommand \href [0]{\begingroup \@sanitize@url \@href}%
\providecommand \@href[1]{\@@startlink{#1}\@@href}%
\providecommand \@@href[1]{\endgroup#1\@@endlink}%
\providecommand \@sanitize@url [0]{\catcode `\\12\catcode `\$12\catcode
  `\&12\catcode `\#12\catcode `\^12\catcode `\_12\catcode `\%12\relax}%
\providecommand \@@startlink[1]{}%
\providecommand \@@endlink[0]{}%
\providecommand \url  [0]{\begingroup\@sanitize@url \@url }%
\providecommand \@url [1]{\endgroup\@href {#1}{\urlprefix }}%
\providecommand \urlprefix  [0]{URL }%
\providecommand \Eprint [0]{\href }%
\providecommand \doibase [0]{http://dx.doi.org/}%
\providecommand \selectlanguage [0]{\@gobble}%
\providecommand \bibinfo  [0]{\@secondoftwo}%
\providecommand \bibfield  [0]{\@secondoftwo}%
\providecommand \translation [1]{[#1]}%
\providecommand \BibitemOpen [0]{}%
\providecommand \bibitemStop [0]{}%
\providecommand \bibitemNoStop [0]{.\EOS\space}%
\providecommand \EOS [0]{\spacefactor3000\relax}%
\providecommand \BibitemShut  [1]{\csname bibitem#1\endcsname}%
\let\auto@bib@innerbib\@empty
\bibitem [{\citenamefont {Boninsegni}\ and\ \citenamefont
  {Prokof'ev}(2012)}]{Boninsegni2012-2}%
  \BibitemOpen
  \bibfield  {author} {\bibinfo {author} {\bibfnamefont {M.}~\bibnamefont
  {Boninsegni}}\ and\ \bibinfo {author} {\bibfnamefont {N.~V.}\ \bibnamefont
  {Prokof'ev}},\ }\href {\doibase 10.1103/RevModPhys.84.759} {\bibfield
  {journal} {\bibinfo  {journal} {Rev. Mod. Phys.}\ }\textbf {\bibinfo {volume}
  {84}},\ \bibinfo {pages} {759} (\bibinfo {year} {2012})}\BibitemShut
  {NoStop}%
\bibitem [{\citenamefont {Nandkishore}\ and\ \citenamefont
  {Huse}(2015)}]{Nandkishore2015}%
  \BibitemOpen
  \bibfield  {author} {\bibinfo {author} {\bibfnamefont {R.}~\bibnamefont
  {Nandkishore}}\ and\ \bibinfo {author} {\bibfnamefont {D.~A.}\ \bibnamefont
  {Huse}},\ }\href {\doibase 10.1146/annurev-conmatphys-031214-014726}
  {\bibfield  {journal} {\bibinfo  {journal} {Annual Review of Condensed Matter
  Physics}\ }\textbf {\bibinfo {volume} {6}},\ \bibinfo {pages} {15} (\bibinfo
  {year} {2015})}\BibitemShut {NoStop}%
\bibitem [{\citenamefont {Qi}\ and\ \citenamefont {S.-C.}(2010)}]{Qi2010}%
  \BibitemOpen
  \bibfield  {author} {\bibinfo {author} {\bibfnamefont {X.-L.}\ \bibnamefont
  {Qi}}\ and\ \bibinfo {author} {\bibfnamefont {Z.}~\bibnamefont {S.-C.}},\
  }\href {\doibase 10.1063/1.3293411} {\bibfield  {journal} {\bibinfo
  {journal} {Physics Today}\ }\textbf {\bibinfo {volume} {63}},\ \bibinfo
  {pages} {33} (\bibinfo {year} {2010})}\BibitemShut {NoStop}%
\bibitem [{\citenamefont {Edwards}\ and\ \citenamefont
  {Anderson}(1975)}]{Edwards1975}%
  \BibitemOpen
  \bibfield  {author} {\bibinfo {author} {\bibfnamefont {S.~F.}\ \bibnamefont
  {Edwards}}\ and\ \bibinfo {author} {\bibfnamefont {P.~W.}\ \bibnamefont
  {Anderson}},\ }\href {http://stacks.iop.org/0305-4608/5/i=5/a=017} {\bibfield
   {journal} {\bibinfo  {journal} {Journal of Physics F: Metal Physics}\
  }\textbf {\bibinfo {volume} {5}},\ \bibinfo {pages} {965} (\bibinfo {year}
  {1975})}\BibitemShut {NoStop}%
\bibitem [{\citenamefont {Binder}\ and\ \citenamefont
  {Young}(1986)}]{Binder1986}%
  \BibitemOpen
  \bibfield  {author} {\bibinfo {author} {\bibfnamefont {K.}~\bibnamefont
  {Binder}}\ and\ \bibinfo {author} {\bibfnamefont {A.~P.}\ \bibnamefont
  {Young}},\ }\href {\doibase 10.1103/RevModPhys.58.801} {\bibfield  {journal}
  {\bibinfo  {journal} {Rev. Mod. Phys.}\ }\textbf {\bibinfo {volume} {58}},\
  \bibinfo {pages} {801} (\bibinfo {year} {1986})}\BibitemShut {NoStop}%
\bibitem [{\citenamefont {Mladek}\ \emph {et~al.}(2006)\citenamefont {Mladek},
  \citenamefont {Gottwald}, \citenamefont {Kahl}, \citenamefont {Neumann},\
  and\ \citenamefont {Likos}}]{Mladek2006}%
  \BibitemOpen
  \bibfield  {author} {\bibinfo {author} {\bibfnamefont {B.~M.}\ \bibnamefont
  {Mladek}}, \bibinfo {author} {\bibfnamefont {D.}~\bibnamefont {Gottwald}},
  \bibinfo {author} {\bibfnamefont {G.}~\bibnamefont {Kahl}}, \bibinfo {author}
  {\bibfnamefont {M.}~\bibnamefont {Neumann}}, \ and\ \bibinfo {author}
  {\bibfnamefont {C.~N.}\ \bibnamefont {Likos}},\ }\href {\doibase
  10.1103/PhysRevLett.96.045701} {\bibfield  {journal} {\bibinfo  {journal}
  {Phys. Rev. Lett.}\ }\textbf {\bibinfo {volume} {96}},\ \bibinfo {pages}
  {045701} (\bibinfo {year} {2006})}\BibitemShut {NoStop}%
\bibitem [{\citenamefont {Lenz}\ \emph {et~al.}(2012)\citenamefont {Lenz},
  \citenamefont {Blaak}, \citenamefont {Likos},\ and\ \citenamefont
  {Mladek}}]{Lenz2012}%
  \BibitemOpen
  \bibfield  {author} {\bibinfo {author} {\bibfnamefont {D.~A.}\ \bibnamefont
  {Lenz}}, \bibinfo {author} {\bibfnamefont {R.}~\bibnamefont {Blaak}},
  \bibinfo {author} {\bibfnamefont {C.~N.}\ \bibnamefont {Likos}}, \ and\
  \bibinfo {author} {\bibfnamefont {B.~M.}\ \bibnamefont {Mladek}},\ }\href
  {\doibase 10.1103/PhysRevLett.109.228301} {\bibfield  {journal} {\bibinfo
  {journal} {Phys. Rev. Lett.}\ }\textbf {\bibinfo {volume} {109}},\ \bibinfo
  {pages} {228301} (\bibinfo {year} {2012})}\BibitemShut {NoStop}%
\bibitem [{\citenamefont {Sciortino}(2013)}]{Sciortino2013}%
  \BibitemOpen
  \bibfield  {author} {\bibinfo {author} {\bibfnamefont {E.}~\bibnamefont
  {Sciortino}, \bibfnamefont {Francesco;~Zaccarelli}},\ }\href {\doibase
  10.1038/493030a} {\bibfield  {journal} {\bibinfo  {journal} {Nature}\
  }\textbf {\bibinfo {volume} {493}},\ \bibinfo {pages} {30} (\bibinfo {year}
  {2013})}\BibitemShut {NoStop}%
\bibitem [{\citenamefont {Saffman}\ \emph {et~al.}(2010)\citenamefont
  {Saffman}, \citenamefont {Walker},\ and\ \citenamefont
  {M{\o}lmer}}]{Saffman2010}%
  \BibitemOpen
  \bibfield  {author} {\bibinfo {author} {\bibfnamefont {M.}~\bibnamefont
  {Saffman}}, \bibinfo {author} {\bibfnamefont {T.}~\bibnamefont {Walker}}, \
  and\ \bibinfo {author} {\bibfnamefont {K.}~\bibnamefont {M{\o}lmer}},\ }\href
  {\doibase 10.1103/RevModPhys.82.2313} {\bibfield  {journal} {\bibinfo
  {journal} {Rev. Mod. Phys.}\ }\textbf {\bibinfo {volume} {82}},\ \bibinfo
  {pages} {2313} (\bibinfo {year} {2010})}\BibitemShut {NoStop}%
\bibitem [{\citenamefont {Jau}\ \emph {et~al.}(2015)\citenamefont {Jau},
  \citenamefont {Hankin}, \citenamefont {Keating}, \citenamefont {Deutsch},\
  and\ \citenamefont {Biedermann}}]{Jau2015}%
  \BibitemOpen
  \bibfield  {author} {\bibinfo {author} {\bibfnamefont {Y.-Y.}\ \bibnamefont
  {Jau}}, \bibinfo {author} {\bibfnamefont {A.~M.}\ \bibnamefont {Hankin}},
  \bibinfo {author} {\bibfnamefont {T.}~\bibnamefont {Keating}}, \bibinfo
  {author} {\bibfnamefont {I.~H.}\ \bibnamefont {Deutsch}}, \ and\ \bibinfo
  {author} {\bibfnamefont {G.~W.}\ \bibnamefont {Biedermann}},\ }\href
  {\doibase 10.1038/nphys3487} {\bibfield  {journal} {\bibinfo  {journal}
  {Nature Physics}\ }\textbf {\bibinfo {volume} {12}},\ \bibinfo {pages} {71}
  (\bibinfo {year} {2015})}\BibitemShut {NoStop}%
\bibitem [{\citenamefont {Zeiher}\ \emph {et~al.}(2016)\citenamefont {Zeiher},
  \citenamefont {van Bijnen}, \citenamefont {Schau{\ss}}, \citenamefont {Hild},
  \citenamefont {yoon Choi}, \citenamefont {Pohl}, \citenamefont {Bloch},\ and\
  \citenamefont {Gross}}]{Zeiher2016}%
  \BibitemOpen
  \bibfield  {author} {\bibinfo {author} {\bibfnamefont {J.}~\bibnamefont
  {Zeiher}}, \bibinfo {author} {\bibfnamefont {R.}~\bibnamefont {van Bijnen}},
  \bibinfo {author} {\bibfnamefont {P.}~\bibnamefont {Schau{\ss}}}, \bibinfo
  {author} {\bibfnamefont {S.}~\bibnamefont {Hild}}, \bibinfo {author}
  {\bibfnamefont {J.}~\bibnamefont {yoon Choi}}, \bibinfo {author}
  {\bibfnamefont {T.}~\bibnamefont {Pohl}}, \bibinfo {author} {\bibfnamefont
  {I.}~\bibnamefont {Bloch}}, \ and\ \bibinfo {author} {\bibfnamefont
  {C.}~\bibnamefont {Gross}},\ }\href {\doibase 10.1038/nphys3835} {\bibfield
  {journal} {\bibinfo  {journal} {Nature Physics}\ }\textbf {\bibinfo {volume}
  {12}},\ \bibinfo {pages} {1095} (\bibinfo {year} {2016})}\BibitemShut
  {NoStop}%
\bibitem [{\citenamefont {Lahaye}\ \emph {et~al.}(2009)\citenamefont {Lahaye},
  \citenamefont {Menotti}, \citenamefont {Santos}, \citenamefont {Lewenstein},\
  and\ \citenamefont {Pfau}}]{Lahaye2009}%
  \BibitemOpen
  \bibfield  {author} {\bibinfo {author} {\bibfnamefont {T.}~\bibnamefont
  {Lahaye}}, \bibinfo {author} {\bibfnamefont {C.}~\bibnamefont {Menotti}},
  \bibinfo {author} {\bibfnamefont {L.}~\bibnamefont {Santos}}, \bibinfo
  {author} {\bibfnamefont {M.}~\bibnamefont {Lewenstein}}, \ and\ \bibinfo
  {author} {\bibfnamefont {T.}~\bibnamefont {Pfau}},\ }\href
  {http://arxiv.org/abs/0905.0386} {\bibfield  {journal} {\bibinfo  {journal}
  {Rep. Prog. Phys.}\ }\textbf {\bibinfo {volume} {72}},\ \bibinfo {pages}
  {126401} (\bibinfo {year} {2009})},\ \Eprint {http://arxiv.org/abs/0905.0386}
  {0905.0386} \BibitemShut {NoStop}%
\bibitem [{\citenamefont {L{\"o}w}\ \emph {et~al.}(2012)\citenamefont
  {L{\"o}w}, \citenamefont {Weimer}, \citenamefont {Nipper}, \citenamefont
  {Balewski}, \citenamefont {Butscher}, \citenamefont {B{\"u}chler},\ and\
  \citenamefont {Pfau}}]{Low:2012aa}%
  \BibitemOpen
  \bibfield  {author} {\bibinfo {author} {\bibfnamefont {R.}~\bibnamefont
  {L{\"o}w}}, \bibinfo {author} {\bibfnamefont {H.}~\bibnamefont {Weimer}},
  \bibinfo {author} {\bibfnamefont {J.}~\bibnamefont {Nipper}}, \bibinfo
  {author} {\bibfnamefont {J.~B.}\ \bibnamefont {Balewski}}, \bibinfo {author}
  {\bibfnamefont {B.}~\bibnamefont {Butscher}}, \bibinfo {author}
  {\bibfnamefont {H.~P.}\ \bibnamefont {B{\"u}chler}}, \ and\ \bibinfo {author}
  {\bibfnamefont {T.}~\bibnamefont {Pfau}},\ }\href@noop {} {\bibfield
  {journal} {\bibinfo  {journal} {J. Phys. B}\ }\textbf {\bibinfo {volume}
  {45}},\ \bibinfo {pages} {113001} (\bibinfo {year} {2012})}\BibitemShut
  {NoStop}%
\bibitem [{\citenamefont {Bernien}\ \emph {et~al.}(2017)\citenamefont
  {Bernien}, \citenamefont {Schwartz}, \citenamefont {Keesling}, \citenamefont
  {Levine}, \citenamefont {Omran}, \citenamefont {Pichler}, \citenamefont
  {Choi}, \citenamefont {Zibrov}, \citenamefont {Endres}, \citenamefont
  {Greiner}, \citenamefont {Vuleti{\'c}},\ and\ \citenamefont
  {Lukin}}]{Bernien:2017aa}%
  \BibitemOpen
  \bibfield  {author} {\bibinfo {author} {\bibfnamefont {H.}~\bibnamefont
  {Bernien}}, \bibinfo {author} {\bibfnamefont {S.}~\bibnamefont {Schwartz}},
  \bibinfo {author} {\bibfnamefont {A.}~\bibnamefont {Keesling}}, \bibinfo
  {author} {\bibfnamefont {H.}~\bibnamefont {Levine}}, \bibinfo {author}
  {\bibfnamefont {A.}~\bibnamefont {Omran}}, \bibinfo {author} {\bibfnamefont
  {H.}~\bibnamefont {Pichler}}, \bibinfo {author} {\bibfnamefont
  {S.}~\bibnamefont {Choi}}, \bibinfo {author} {\bibfnamefont {A.~S.}\
  \bibnamefont {Zibrov}}, \bibinfo {author} {\bibfnamefont {M.}~\bibnamefont
  {Endres}}, \bibinfo {author} {\bibfnamefont {M.}~\bibnamefont {Greiner}},
  \bibinfo {author} {\bibfnamefont {V.}~\bibnamefont {Vuleti{\'c}}}, \ and\
  \bibinfo {author} {\bibfnamefont {M.~D.}\ \bibnamefont {Lukin}},\ }\href@noop
  {} {\bibfield  {journal} {\bibinfo  {journal} {Nature}\ }\textbf {\bibinfo
  {volume} {551}},\ \bibinfo {pages} {579} (\bibinfo {year}
  {2017})}\BibitemShut {NoStop}%
\bibitem [{\citenamefont {de~Léséleuc}\ \emph {et~al.}()\citenamefont
  {de~Léséleuc}, \citenamefont {Lienhard}, \citenamefont {Scholl},
  \citenamefont {Barredo}, \citenamefont {Weber}, \citenamefont {Lang},
  \citenamefont {Büchler}, \citenamefont {Lahaye},\ and\ \citenamefont
  {Browaeys}}]{Leseleuc:aa}%
  \BibitemOpen
  \bibfield  {author} {\bibinfo {author} {\bibfnamefont {S.}~\bibnamefont
  {de~Léséleuc}}, \bibinfo {author} {\bibfnamefont {V.}~\bibnamefont
  {Lienhard}}, \bibinfo {author} {\bibfnamefont {P.}~\bibnamefont {Scholl}},
  \bibinfo {author} {\bibfnamefont {D.}~\bibnamefont {Barredo}}, \bibinfo
  {author} {\bibfnamefont {S.}~\bibnamefont {Weber}}, \bibinfo {author}
  {\bibfnamefont {N.}~\bibnamefont {Lang}}, \bibinfo {author} {\bibfnamefont
  {H.~P.}\ \bibnamefont {Büchler}}, \bibinfo {author} {\bibfnamefont
  {T.}~\bibnamefont {Lahaye}}, \ and\ \bibinfo {author} {\bibfnamefont
  {A.}~\bibnamefont {Browaeys}},\ }\href@noop {} {}\Eprint
  {http://arxiv.org/abs/arXiv:1810.13286} {arXiv:1810.13286} \BibitemShut
  {NoStop}%
\bibitem [{\citenamefont {D\'{\i}az-M\'endez}\ \emph
  {et~al.}(2015)\citenamefont {D\'{\i}az-M\'endez}, \citenamefont {Mezzacapo},
  \citenamefont {Cinti}, \citenamefont {Lechner},\ and\ \citenamefont
  {Pupillo}}]{Diaz-Mendez2015}%
  \BibitemOpen
  \bibfield  {author} {\bibinfo {author} {\bibfnamefont {R.}~\bibnamefont
  {D\'{\i}az-M\'endez}}, \bibinfo {author} {\bibfnamefont {F.}~\bibnamefont
  {Mezzacapo}}, \bibinfo {author} {\bibfnamefont {F.}~\bibnamefont {Cinti}},
  \bibinfo {author} {\bibfnamefont {W.}~\bibnamefont {Lechner}}, \ and\
  \bibinfo {author} {\bibfnamefont {G.}~\bibnamefont {Pupillo}},\ }\href
  {\doibase 10.1103/PhysRevE.92.052307} {\bibfield  {journal} {\bibinfo
  {journal} {Phys. Rev. E}\ }\textbf {\bibinfo {volume} {92}},\ \bibinfo
  {pages} {052307} (\bibinfo {year} {2015})}\BibitemShut {NoStop}%
\bibitem [{\citenamefont {D\'{\i}az-M\'endez}\ \emph
  {et~al.}(2017)\citenamefont {D\'{\i}az-M\'endez}, \citenamefont {Mezzacapo},
  \citenamefont {Lechner}, \citenamefont {Cinti}, \citenamefont {Babaev},\ and\
  \citenamefont {Pupillo}}]{Diaz-Mendez2017}%
  \BibitemOpen
  \bibfield  {author} {\bibinfo {author} {\bibfnamefont {R.}~\bibnamefont
  {D\'{\i}az-M\'endez}}, \bibinfo {author} {\bibfnamefont {F.}~\bibnamefont
  {Mezzacapo}}, \bibinfo {author} {\bibfnamefont {W.}~\bibnamefont {Lechner}},
  \bibinfo {author} {\bibfnamefont {F.}~\bibnamefont {Cinti}}, \bibinfo
  {author} {\bibfnamefont {E.}~\bibnamefont {Babaev}}, \ and\ \bibinfo {author}
  {\bibfnamefont {G.}~\bibnamefont {Pupillo}},\ }\href {\doibase
  10.1103/PhysRevLett.118.067001} {\bibfield  {journal} {\bibinfo  {journal}
  {Phys. Rev. Lett.}\ }\textbf {\bibinfo {volume} {118}},\ \bibinfo {pages}
  {067001} (\bibinfo {year} {2017})}\BibitemShut {NoStop}%
\bibitem [{\citenamefont {Diaz-Mendez}\ \emph {et~al.}(2019)\citenamefont
  {Diaz-Mendez}, \citenamefont {Pupillo}, \citenamefont {Mezzacapo},
  \citenamefont {Wallin}, \citenamefont {Lidmar},\ and\ \citenamefont
  {Babaev}}]{Diaz-Mendez2019}%
  \BibitemOpen
  \bibfield  {author} {\bibinfo {author} {\bibfnamefont {R.}~\bibnamefont
  {Diaz-Mendez}}, \bibinfo {author} {\bibfnamefont {G.}~\bibnamefont
  {Pupillo}}, \bibinfo {author} {\bibfnamefont {F.}~\bibnamefont {Mezzacapo}},
  \bibinfo {author} {\bibfnamefont {M.}~\bibnamefont {Wallin}}, \bibinfo
  {author} {\bibfnamefont {J.}~\bibnamefont {Lidmar}}, \ and\ \bibinfo {author}
  {\bibfnamefont {E.}~\bibnamefont {Babaev}},\ }\href {\doibase
  10.1039/C8SM01738G} {\bibfield  {journal} {\bibinfo  {journal} {Soft Matter}\
  }\textbf {\bibinfo {volume} {15}},\ \bibinfo {pages} {355} (\bibinfo {year}
  {2019})}\BibitemShut {NoStop}%
\bibitem [{\citenamefont {Mattioli}\ \emph {et~al.}(2013)\citenamefont
  {Mattioli}, \citenamefont {Dalmonte}, \citenamefont {Lechner},\ and\
  \citenamefont {Pupillo}}]{Mattioli2013}%
  \BibitemOpen
  \bibfield  {author} {\bibinfo {author} {\bibfnamefont {M.}~\bibnamefont
  {Mattioli}}, \bibinfo {author} {\bibfnamefont {M.}~\bibnamefont {Dalmonte}},
  \bibinfo {author} {\bibfnamefont {W.}~\bibnamefont {Lechner}}, \ and\
  \bibinfo {author} {\bibfnamefont {G.}~\bibnamefont {Pupillo}},\ }\href
  {\doibase 10.1103/PhysRevLett.111.165302} {\bibfield  {journal} {\bibinfo
  {journal} {Phys. Rev. Lett.}\ }\textbf {\bibinfo {volume} {111}},\ \bibinfo
  {pages} {165302} (\bibinfo {year} {2013})}\BibitemShut {NoStop}%
\bibitem [{\citenamefont {Dalmonte}\ \emph {et~al.}(2015)\citenamefont
  {Dalmonte}, \citenamefont {Lechner}, \citenamefont {Cai}, \citenamefont
  {Mattioli}, \citenamefont {L{\"a}uchli},\ and\ \citenamefont
  {Pupillo}}]{Dalmonte:2015aa}%
  \BibitemOpen
  \bibfield  {author} {\bibinfo {author} {\bibfnamefont {M.}~\bibnamefont
  {Dalmonte}}, \bibinfo {author} {\bibfnamefont {W.}~\bibnamefont {Lechner}},
  \bibinfo {author} {\bibfnamefont {Z.}~\bibnamefont {Cai}}, \bibinfo {author}
  {\bibfnamefont {M.}~\bibnamefont {Mattioli}}, \bibinfo {author}
  {\bibfnamefont {A.~M.}\ \bibnamefont {L{\"a}uchli}}, \ and\ \bibinfo {author}
  {\bibfnamefont {G.}~\bibnamefont {Pupillo}},\ }\href@noop {} {\bibfield
  {journal} {\bibinfo  {journal} {Phys. Rev. B}\ }\textbf {\bibinfo {volume}
  {92}},\ \bibinfo {pages} {045106} (\bibinfo {year} {2015})}\BibitemShut
  {NoStop}%
\bibitem [{\citenamefont {Motta}\ \emph {et~al.}(2016)\citenamefont {Motta},
  \citenamefont {Vitali}, \citenamefont {Rossi}, \citenamefont {Galli},\ and\
  \citenamefont {Bertaina}}]{Motta:2016aa}%
  \BibitemOpen
  \bibfield  {author} {\bibinfo {author} {\bibfnamefont {M.}~\bibnamefont
  {Motta}}, \bibinfo {author} {\bibfnamefont {E.}~\bibnamefont {Vitali}},
  \bibinfo {author} {\bibfnamefont {M.}~\bibnamefont {Rossi}}, \bibinfo
  {author} {\bibfnamefont {D.~E.}\ \bibnamefont {Galli}}, \ and\ \bibinfo
  {author} {\bibfnamefont {G.}~\bibnamefont {Bertaina}},\ }\href@noop {}
  {\bibfield  {journal} {\bibinfo  {journal} {Phys. Rev. A}\ }\textbf {\bibinfo
  {volume} {94}},\ \bibinfo {pages} {043627} (\bibinfo {year}
  {2016})}\BibitemShut {NoStop}%
\bibitem [{\citenamefont {Teruzzi}\ \emph {et~al.}(2017)\citenamefont
  {Teruzzi}, \citenamefont {Galli},\ and\ \citenamefont
  {Bertaina}}]{Teruzzi:aa}%
  \BibitemOpen
  \bibfield  {author} {\bibinfo {author} {\bibfnamefont {M.}~\bibnamefont
  {Teruzzi}}, \bibinfo {author} {\bibfnamefont {D.~E.}\ \bibnamefont {Galli}},
  \ and\ \bibinfo {author} {\bibfnamefont {G.}~\bibnamefont {Bertaina}},\
  }\href {\doibase 10.1007/s10909-016-1736-0} {\bibfield  {journal} {\bibinfo
  {journal} {Journal of Low Temperature Physics}\ }\textbf {\bibinfo {volume}
  {187}},\ \bibinfo {pages} {719} (\bibinfo {year} {2017})}\BibitemShut
  {NoStop}%
\bibitem [{\citenamefont {Henkel}\ \emph {et~al.}(2010)\citenamefont {Henkel},
  \citenamefont {Nath},\ and\ \citenamefont {Pohl}}]{Henkel2010}%
  \BibitemOpen
  \bibfield  {author} {\bibinfo {author} {\bibfnamefont {N.}~\bibnamefont
  {Henkel}}, \bibinfo {author} {\bibfnamefont {R.}~\bibnamefont {Nath}}, \ and\
  \bibinfo {author} {\bibfnamefont {T.}~\bibnamefont {Pohl}},\ }\href {\doibase
  10.1103/PhysRevLett.104.195302} {\bibfield  {journal} {\bibinfo  {journal}
  {Phys. Rev. Lett.}\ }\textbf {\bibinfo {volume} {104}},\ \bibinfo {pages}
  {195302} (\bibinfo {year} {2010})}\BibitemShut {NoStop}%
\bibitem [{\citenamefont {Henkel}\ \emph {et~al.}(2012)\citenamefont {Henkel},
  \citenamefont {Cinti}, \citenamefont {Jain}, \citenamefont {Pupillo},\ and\
  \citenamefont {Pohl}}]{Henkel2012}%
  \BibitemOpen
  \bibfield  {author} {\bibinfo {author} {\bibfnamefont {N.}~\bibnamefont
  {Henkel}}, \bibinfo {author} {\bibfnamefont {F.}~\bibnamefont {Cinti}},
  \bibinfo {author} {\bibfnamefont {P.}~\bibnamefont {Jain}}, \bibinfo {author}
  {\bibfnamefont {G.}~\bibnamefont {Pupillo}}, \ and\ \bibinfo {author}
  {\bibfnamefont {T.}~\bibnamefont {Pohl}},\ }\href {\doibase
  10.1103/PhysRevLett.108.265301} {\bibfield  {journal} {\bibinfo  {journal}
  {Phys. Rev. Lett.}\ }\textbf {\bibinfo {volume} {108}},\ \bibinfo {pages}
  {265301} (\bibinfo {year} {2012})}\BibitemShut {NoStop}%
\bibitem [{\citenamefont {Cinti}\ \emph {et~al.}(2010)\citenamefont {Cinti},
  \citenamefont {Jain}, \citenamefont {Boninsegni}, \citenamefont {Micheli},
  \citenamefont {Zoller},\ and\ \citenamefont {Pupillo}}]{Cinti2010}%
  \BibitemOpen
  \bibfield  {author} {\bibinfo {author} {\bibfnamefont {F.}~\bibnamefont
  {Cinti}}, \bibinfo {author} {\bibfnamefont {P.}~\bibnamefont {Jain}},
  \bibinfo {author} {\bibfnamefont {M.}~\bibnamefont {Boninsegni}}, \bibinfo
  {author} {\bibfnamefont {A.}~\bibnamefont {Micheli}}, \bibinfo {author}
  {\bibfnamefont {P.}~\bibnamefont {Zoller}}, \ and\ \bibinfo {author}
  {\bibfnamefont {G.}~\bibnamefont {Pupillo}},\ }\href {\doibase
  10.1103/PhysRevLett.105.135301} {\bibfield  {journal} {\bibinfo  {journal}
  {Phys. Rev. Lett.}\ }\textbf {\bibinfo {volume} {105}},\ \bibinfo {pages}
  {135301} (\bibinfo {year} {2010})}\BibitemShut {NoStop}%
\bibitem [{\citenamefont {Cinti}\ \emph {et~al.}(2014)\citenamefont {Cinti},
  \citenamefont {Macrì}, \citenamefont {Lechner}, \citenamefont {Pupillo},\
  and\ \citenamefont {Pohl}}]{Cinti2014}%
  \BibitemOpen
  \bibfield  {author} {\bibinfo {author} {\bibfnamefont {F.}~\bibnamefont
  {Cinti}}, \bibinfo {author} {\bibfnamefont {T.}~\bibnamefont {Macrì}},
  \bibinfo {author} {\bibfnamefont {W.}~\bibnamefont {Lechner}}, \bibinfo
  {author} {\bibfnamefont {G.}~\bibnamefont {Pupillo}}, \ and\ \bibinfo
  {author} {\bibfnamefont {T.}~\bibnamefont {Pohl}},\ }\href {\doibase
  10.1038/ncomms4235} {\bibfield  {journal} {\bibinfo  {journal} {Nature
  Communications}\ }\textbf {\bibinfo {volume} {5}},\ \bibinfo {pages} {1}
  (\bibinfo {year} {2014})}\BibitemShut {NoStop}%
\bibitem [{\citenamefont {Angelone}\ \emph {et~al.}(2016)\citenamefont
  {Angelone}, \citenamefont {Mezzacapo},\ and\ \citenamefont
  {Pupillo}}]{Angelone2016}%
  \BibitemOpen
  \bibfield  {author} {\bibinfo {author} {\bibfnamefont {A.}~\bibnamefont
  {Angelone}}, \bibinfo {author} {\bibfnamefont {F.}~\bibnamefont {Mezzacapo}},
  \ and\ \bibinfo {author} {\bibfnamefont {G.}~\bibnamefont {Pupillo}},\ }\href
  {\doibase 10.1103/PhysRevLett.116.135303} {\bibfield  {journal} {\bibinfo
  {journal} {Phys. Rev. Lett.}\ }\textbf {\bibinfo {volume} {116}},\ \bibinfo
  {pages} {135303} (\bibinfo {year} {2016})}\BibitemShut {NoStop}%
\bibitem [{\citenamefont {Masella}\ \emph {et~al.}(2019)\citenamefont
  {Masella}, \citenamefont {Angelone}, \citenamefont {Mezzacapo}, \citenamefont
  {Pupillo},\ and\ \citenamefont {Prokof'ev}}]{Masella2019}%
  \BibitemOpen
  \bibfield  {author} {\bibinfo {author} {\bibfnamefont {G.}~\bibnamefont
  {Masella}}, \bibinfo {author} {\bibfnamefont {A.}~\bibnamefont {Angelone}},
  \bibinfo {author} {\bibfnamefont {F.}~\bibnamefont {Mezzacapo}}, \bibinfo
  {author} {\bibfnamefont {G.}~\bibnamefont {Pupillo}}, \ and\ \bibinfo
  {author} {\bibfnamefont {N.~V.}\ \bibnamefont {Prokof'ev}},\ }\href {\doibase
  10.1103/PhysRevLett.123.045301} {\bibfield  {journal} {\bibinfo  {journal}
  {Phys. Rev. Lett.}\ }\textbf {\bibinfo {volume} {123}},\ \bibinfo {pages}
  {045301} (\bibinfo {year} {2019})}\BibitemShut {NoStop}%
\bibitem [{\citenamefont {Menotti}\ \emph {et~al.}(2007)\citenamefont
  {Menotti}, \citenamefont {Trefzger},\ and\ \citenamefont
  {Lewenstein}}]{Menotti2007}%
  \BibitemOpen
  \bibfield  {author} {\bibinfo {author} {\bibfnamefont {C.}~\bibnamefont
  {Menotti}}, \bibinfo {author} {\bibfnamefont {C.}~\bibnamefont {Trefzger}}, \
  and\ \bibinfo {author} {\bibfnamefont {M.}~\bibnamefont {Lewenstein}},\
  }\href {\doibase 10.1103/PhysRevLett.98.235301} {\bibfield  {journal}
  {\bibinfo  {journal} {Phys. Rev. Lett.}\ }\textbf {\bibinfo {volume} {98}},\
  \bibinfo {pages} {235301} (\bibinfo {year} {2007})}\BibitemShut {NoStop}%
\bibitem [{\citenamefont {Trefzger}\ \emph {et~al.}(2008)\citenamefont
  {Trefzger}, \citenamefont {Menotti},\ and\ \citenamefont
  {Lewenstein}}]{Trefzger2008}%
  \BibitemOpen
  \bibfield  {author} {\bibinfo {author} {\bibfnamefont {C.}~\bibnamefont
  {Trefzger}}, \bibinfo {author} {\bibfnamefont {C.}~\bibnamefont {Menotti}}, \
  and\ \bibinfo {author} {\bibfnamefont {M.}~\bibnamefont {Lewenstein}},\
  }\href {\doibase 10.1103/PhysRevA.78.043604} {\bibfield  {journal} {\bibinfo
  {journal} {Phys. Rev. A}\ }\textbf {\bibinfo {volume} {78}},\ \bibinfo
  {pages} {043604} (\bibinfo {year} {2008})}\BibitemShut {NoStop}%
\bibitem [{\citenamefont {Bloch}\ \emph {et~al.}(2012)\citenamefont {Bloch},
  \citenamefont {Dalibard},\ and\ \citenamefont {Nascimbene}}]{Bloch2012}%
  \BibitemOpen
  \bibfield  {author} {\bibinfo {author} {\bibfnamefont {I.}~\bibnamefont
  {Bloch}}, \bibinfo {author} {\bibfnamefont {J.}~\bibnamefont {Dalibard}}, \
  and\ \bibinfo {author} {\bibfnamefont {S.}~\bibnamefont {Nascimbene}},\
  }\href@noop {} {\bibfield  {journal} {\bibinfo  {journal} {Nature Physics}\
  }\textbf {\bibinfo {volume} {8}},\ \bibinfo {pages} {267} (\bibinfo {year}
  {2012})}\BibitemShut {NoStop}%
\bibitem [{\citenamefont {Schmid}\ \emph {et~al.}(2002)\citenamefont {Schmid},
  \citenamefont {Todo}, \citenamefont {Troyer},\ and\ \citenamefont
  {Dorneich}}]{Schmid2002}%
  \BibitemOpen
  \bibfield  {author} {\bibinfo {author} {\bibfnamefont {G.}~\bibnamefont
  {Schmid}}, \bibinfo {author} {\bibfnamefont {S.}~\bibnamefont {Todo}},
  \bibinfo {author} {\bibfnamefont {M.}~\bibnamefont {Troyer}}, \ and\ \bibinfo
  {author} {\bibfnamefont {A.}~\bibnamefont {Dorneich}},\ }\href {\doibase
  10.1103/PhysRevLett.88.167208} {\bibfield  {journal} {\bibinfo  {journal}
  {Phys. Rev. Lett.}\ }\textbf {\bibinfo {volume} {88}},\ \bibinfo {pages}
  {167208} (\bibinfo {year} {2002})}\BibitemShut {NoStop}%
\bibitem [{\citenamefont {Batrouni}\ and\ \citenamefont
  {Scalettar}(2000)}]{Batrouni2000}%
  \BibitemOpen
  \bibfield  {author} {\bibinfo {author} {\bibfnamefont {G.~G.}\ \bibnamefont
  {Batrouni}}\ and\ \bibinfo {author} {\bibfnamefont {R.~T.}\ \bibnamefont
  {Scalettar}},\ }\href {\doibase 10.1103/PhysRevLett.84.1599} {\bibfield
  {journal} {\bibinfo  {journal} {Phys. Rev. Lett.}\ }\textbf {\bibinfo
  {volume} {84}},\ \bibinfo {pages} {1599} (\bibinfo {year}
  {2000})}\BibitemShut {NoStop}%
\bibitem [{\citenamefont {H\'ebert}\ \emph {et~al.}(2001)\citenamefont
  {H\'ebert}, \citenamefont {Batrouni}, \citenamefont {Scalettar},
  \citenamefont {Schmid}, \citenamefont {Troyer},\ and\ \citenamefont
  {Dorneich}}]{Hebert2002}%
  \BibitemOpen
  \bibfield  {author} {\bibinfo {author} {\bibfnamefont {F.}~\bibnamefont
  {H\'ebert}}, \bibinfo {author} {\bibfnamefont {G.~G.}\ \bibnamefont
  {Batrouni}}, \bibinfo {author} {\bibfnamefont {R.~T.}\ \bibnamefont
  {Scalettar}}, \bibinfo {author} {\bibfnamefont {G.}~\bibnamefont {Schmid}},
  \bibinfo {author} {\bibfnamefont {M.}~\bibnamefont {Troyer}}, \ and\ \bibinfo
  {author} {\bibfnamefont {A.}~\bibnamefont {Dorneich}},\ }\href {\doibase
  10.1103/PhysRevB.65.014513} {\bibfield  {journal} {\bibinfo  {journal} {Phys.
  Rev. B}\ }\textbf {\bibinfo {volume} {65}},\ \bibinfo {pages} {014513}
  (\bibinfo {year} {2001})}\BibitemShut {NoStop}%
\bibitem [{\citenamefont {Prokof'ev}\ \emph {et~al.}(1998)\citenamefont
  {Prokof'ev}, \citenamefont {Svistunov},\ and\ \citenamefont
  {Tupitsyn}}]{Prokofev1998-1}%
  \BibitemOpen
  \bibfield  {author} {\bibinfo {author} {\bibfnamefont {N.~V.}\ \bibnamefont
  {Prokof'ev}}, \bibinfo {author} {\bibfnamefont {B.~V.}\ \bibnamefont
  {Svistunov}}, \ and\ \bibinfo {author} {\bibfnamefont {I.~S.}\ \bibnamefont
  {Tupitsyn}},\ }\href {\doibase 10.1134/1.558661} {\bibfield  {journal}
  {\bibinfo  {journal} {Journal of Experimental and Theoretical Physics}\
  }\textbf {\bibinfo {volume} {87}},\ \bibinfo {pages} {310} (\bibinfo {year}
  {1998})}\BibitemShut {NoStop}%
\bibitem [{que()}]{questionB-ref1}%
  \BibitemOpen
  \href@noop {} {}\bibinfo {howpublished} {It has to be stressed that the
  glassy scenarios that we find at large $V/t$, i.e., where the system is
  essentially classical, have been obtained by some of us in similar models via
  molecular dynamics approaches which give access to real-time
  dynamics~\cite{Diaz-Mendez2017}. This points out how the appearance of stable
  disordered states in our work is not an artifact of the adopted QMC
  algorithm. The same applies for the OOE states at intermediate $V/t$ where
  our algorithm becomes even more efficient. Indeed, for small $V/t$ our quench
  protocol is ineffective, and equilibrium is reached even if annealing is not
  performed.}\BibitemShut {Stop}%
\end{thebibliography}%
\end{document}